\documentclass[openacc]{rstransa}

\titlehead{Research}

\begin{document}

\title{Status of the $\Lambda$CDM theory: supporting evidence and anomalies}

\author{
P. J. E. Peebles$^{1}$}

\address{$^{1}$Joseph Henry Laboratories\\Princeton University\\
Princeton NJ USA}

\subject{Physical Cosmology}


\corres{P. J. . Peebles\\
\email{pjep@princeton.edu}}

\begin{abstract}
The standard $\Lambda$CDM cosmology passes demanding tests that establish it as a good approximation to reality. It is incomplete, with open questions and anomalies, but the same is true of all our physical theories. The anomalies in the standard cosmology might guide us to an even better theory. It has happened before. 
 \end{abstract}


\begin{fmtext}

\end{fmtext}


\maketitle

My thinking about the state of the standard $\Lambda$CDM cosmology is guided by my understanding of the history of how physics grew. Consider the state of fundamental physics a century ago. In 1924 Maxwell's theory of electromagnetism had passed demanding tests in the laboratory and far more tests by the great variety of practical applications, from radio to streetcars. But physicists knew full well  that the theory fails on the scale of atoms. Maybe some were still interested in a possible failure of the inverse square law on small scales, but most were thinking about the many empirical clues to the quantization of energy in matter, as in spectral lines, and in radiation, as in the thermal spectrum and Compton scattering. In the following two years physicists arrived at matrix mechanics and wave mechanics, and in 1927 Dirac wrote down the application of quantum theory to electromagnetism. These steps to Quantum Electrodynamics, QED, with the great advances in technology during World War II, produced the wonderful precision of experimental tests of predictions of QED. In a 1970 review Brodsky and Drell\cite{QED} wrote that
\begin{quotation}
\noindent QED has a very simple conceptual basis and indeed is built by purely
imitative steps. Shortly after the birth of quantum mechanics it was constructed
very simply by applying the ordinary rules of the quantum theory
to the electromagnetic field amplitudes, $E(y, t)$ and $B (y, t)$, whose space-time
development is given by the Maxwell equations. Thus, as had originally
happened to the position and momentum coordinates of a single particle, the
field amplitudes also became operators whose matrix elements are observable.
\end{quotation}

Each great advance in physics has its own story, but Brodsky and Drell present the common theme of preservation of what had been empirically proved to work and upon which a better theory was constructed. Copernicus rationalized the Ptolemaic system by moving the sun to the center of the planetary system, but kept the  Ptolemaic methodology because its predictions were accurate. Einstein replaced Newtonian gravity by the general theory of relativity, with a new vision of spacetime. But we still speak of tides, the rotation of  Foucault's pendulum, and the formation of tornados in Newtonian terms, because they are quite sufficient and a full application of general relativity would be needlessly complicated. QED replaced Maxwell's equations, but electricians still check the continuity of an electric circuit with a multimeter calibrated in the volts, amperes, and ohms of Maxwell's equations. I like the illustration of this point in the Landau and Lifshitz book, The Classical Theory of Fields. In my 1951 edition\cite{LandauLifshitz}, the translation of the 1948 Russian edition, the first two thirds of the book presents thorough analyses of Maxwell's equations. The last third presents Einstein's general theory of relativity. It was generous to assign this much space to general relativity because it had little empirical support; it was there because it is elegant. Now the theory is persuasively established by the demanding tests of predictions it passes. This is how we separate ideas no matter how elegant from what we take to be convincing approximations to reality. 
 
Now we have two well tested approximations to reality. The theory on small scales grew from Maxwell's equations to QED and on in steps that might be termed revolutionary but grew on what works empirically. The theory of reality on large scales has a different kind of origin story, from Einstein's vision of a logically compelling replacement of gravity in a logically compelling universe. This was little more than a vision for the first half century, but now we have serious empirical support for the FLRW $\Lambda$CDM cosmology. The theory is incomplete and we hope a better theory will be found; it has happed before. The history of physics leads us to expect that the better theory will incorporate the empirically successful parts of $\Lambda$CDM, with embellishments. 

\section{Supporting Evidence}\label{sec:evidence}

The central piece of evidence for the standard and accepted relativistic $\Lambda$CDM cosmology is the beautiful agreement between the theory and measurements of the power spectra of the temperature and polarization of the thermal cosmic background radiation, the CMB. This test is possible because radio interference from Earth and extraterrestrial sources is modest enough to allow precision measurements. Equally important, the computation of the predictions are simple enough---linear perturbation theory--- to allow accuracy of predictions comparable to that of the measurements. 

I recall a physicist who was arguing for a new theory of gravity stating that the fit of the new theory to the CMB observations and the other cosmological tests will only require reworking some numerical integrations. That remains to be seen, but the challenges for a new theory look far from this simple. The $\Lambda$CDM theory successfully predicts the power spectrum of the spatial distribution of luminous galaxies as well as the CMB angular power spectra\cite{SDSS}. Matter and radiation evolved through decoupling in different ways, and the observable results are measured by observations of different phenomena, but they agree with the same theory\cite{PJEP_tests}. That is a serious addition to the challenge for a new and different theory. The fit of the spectra to the measurements is aided by adjustment of the cosmological parameters, but that offers more tests. Recall that the fit of measured abundances of the stable isotopes of hydrogen and helium to their predicted formation as the universe expanded and cooled through redshift $z\sim 10^9$ requires a baryon abundance that agrees with what is required to fit the CMB anisotropy spectrum frozen in at $z\sim 10^3$, and agrees with observations of the baryon abundance in the plasma in clusters of galaxies and with the abundance in intergalactic space at low redshifts probed by Fast Radio Bursts. Another example is the value of Hubble's constant $H_o$ derived from the fit of measurements of how the CMB anisotropy spectra were shaped by what happened at decoupling at redshift $z\sim 10^3$, which is within 10\%\ of the value derived from the distances and redshifts of galaxies observed at redshifts less than unity. These are at best 10\%\ measurements. Instead of precision the long baseline in cosmic expansion makes these and the other multiply constrained parameters a serious challenge to the search for an alternative theory. 

\section{Anomalies}

I remarked that fundamental physics a century ago had a theory of electromagnetism that passed demanding tests, with anomalies. Cosmology today passes demanding tests, with anomalies. You could argue that the anomalies falsified Maxwell's equations, but better put is that they led to improvements. Anomalies in cosmology need not falsify $\Lambda$CDM; we are seeking evidence to guide thinking of what to make of them and how to improve the theory. 

There are differences between the situations now and a century ago. The problem with atoms and radiation in 1924 was far more pronounced than the anomalies in our standard cosmology. On the other hand, physicists a century ago had no reason to doubt Maxwell's equations, apart from those pesky anomalies, while our cosmology depends on questionable assumptions of simplicity. My {\it ad hoc} introduction of dark matter and reintroduction of the cosmological constant in the 1980s were meant to be simple ways to avoid problems with modest empirical evidence. This model for the dark sector remains consistent with far tighter evidence, apart from those pesky anomalies, but many agree that the dark sector is suspiciously simple compared to the rich physics in the visible sector. A half century ago it was natural to assume Gaussian initial departures from homogeneity, another argument from simplicity rather than empirical evidence. Primeval Gaussianity also agrees with precision CMB measurements. General relativity is a shining example of a successful argument from elegant simplicity. But arguments from empirical evidence are more trustworthy.

I offer in this section thoughts about issues I consider interesting and maybe hints to remedies to overly simple ideas in cosmology. A longer review of my thinking is presented in\cite{PJEPanomalies}. Others have other ideas about the interesting issues in cosmology, as in\cite{anomalies_review}. But the following examples illustrate the  state of the subject.

\subsection{The abundance of  lithium}

In Section~\ref{sec:evidence} I celebrated the consistency of the predicted abundances of the stable isotopes of hydrogen and helium with the primeval abundances derived from observations. It requires the assurance of knowledgeable astronomers that the effects of astrophysical process on these abundances are well enough understood; for details I refer you to a knowledgeable astronomer. The standard cosmology also predicts formation of a measurable abundance of the lithium isotope $\hbox{Li}^7$. The prediction is secure, but the effects of the passage of matter through stars where $\hbox{Li}^7$ is readily broken up, and the production of lithium in supernovae, are not understood. The assessment by Matthews et al.\cite{BBNS} is
\begin{quotation}
\noindent The observationally inferred abundance of primordial lithium remains at about a
factor of three below the abundance predicted by standard big bang nucleosynthesis (BBN).
The resolution of this dilemma can be either astrophysical (stars destroy lithium after BBN),
nuclear (reactions destroy lithium during BBN), or cosmological, i.e. new physics beyond the
standard BBN is responsible for destroying lithium.
\end{quotation}
This has been a perplexing problem for the past half century. 

\subsection{The Hubble tension}

In the previous section I also celebrated the close agreement of values of the Hubble parameter $H_o$ found from  two quite different probes, the relation between redshifts and distances of galaxies at redshifts $z\lesssim 1$ and the pattern left in the angular distribution of the CMB by the process of decoupling at redshifts $z\gtrsim 10^3$. I mean it. But the two measures of $H_o$ disagree by 5 to 10\%. We are fortunate that Wendy Freedman is presenting an informed assessment of the anomaly in these proceedings. Maybe the anomaly will be resolved by the discovery of some subtle overlooked observational effect. Or perhaps it will be resolved by the discovery of a better theory that does away with the hypothetical dark matter and dark energy. I doubt that because it would require the discovery of a new way to pass demanding cosmological tests. Then there is the idea of a more interesting dark sector, maybe some variant of Early Dark Energy\cite{dark_energy}.

\subsection{Anisotropic large-scale distributions of extragalactic objects}\label{sec:dipole_anomaly}

Ellis and Baldwin\cite{EllisBaldwin} worked out the local Lorentz transformation from an observer who sees an isotropic large-scale distribution of matter to what is seen by an observer moving relative to this first observer. The computation is similar to the demonstration that an observer moving through a uniform sea of thermal radiation such as the CMB sees an anisotropy in the temperature but with a thermal spectrum in the radiation received in each direction. The computation of the effect for counts of objects as a function of direction is more complicated because it depends on the spectra of the objects and their abundance as a function of apparent magnitude or flux density. But one can deal with this. 

The measured dipole anisotropies in the large-scale distributions of radio galaxies, quasars, and clusters of galaxies tend to be maximum at directions not far from what is predicted by the Lorentz transformation, but most arrive at about twice the predicted size of the anisotropy. Secrest, von Hausegger, Rameez et al.\cite{Sarkar_etal} and Singal\cite{Singal} present assessments of this phenomenon and guides to earlier research. This is good science. But these authors conclude that these findings 
\begin{quotation}
\noindent present a significant challenge to the cosmological principle and, by extension, the standard FLRW cosmological model\cite{Sarkar_etal},
\end{quotation}
and raise
\begin{quotation}
\noindent doubts about the CP \ldots\ the basic foundation of the standard model in modern cosmology\cite{Singal}.
\end{quotation}
This is an unnecessary distraction from a potentially significant anomaly. 

The ``challenge to the cosmological principle'' depends on its definition. These authors take it to be that the universe observed on the largest practical scales is isotropic. But the present standard cosmology assumes the primeval departures from exact homogeneity and isotropy are a stationary random process with a Gaussian power law power spectrum that extends to scales large compared to the present Hubble length. Under this assumption the universe is not isotropic on the largest scales that can be probed. That is, the cosmology need not be challenged by evidence of anisotropy. That depends on a secure demonstration of what the cosmology predicts, whether on not it is consistent with secure observational evidence. 

If the observed large-scale dipole anisotropy in counts of extragalactic objects proved to be inconsistent with the Gaussian adiabatic initial conditions of the standard theory then my first thought would be to explore whether the simple assumption of Gaussian adiabatic initial conditions might have to be adjusted to accommodate the evidence. Would this be possible without an unacceptable effect on the CMB? If so the standard cosmology would have to be adjusted, but I would expect it to be an embellishment. If not the standard cosmology might have be more seriously reconsidered.

\subsection{The extended Local Supercluster}

The trace of non-Gaussianity in the initial conditions for cosmic structure formation suggested in Section~\ref{sec:dipole_anomaly} also is suggested by another curious phenomenon. Many of the galaxies closer than about 10~Mpc are near the flat plane of the de Vaucouleurs Local Supercluster. It has long been known that clusters of galaxies and radio galaxies tend to be near this plane at far greater distances. A recent analysis\cite{cosmic_sheets} of the better data we have now indicates the presence of a significant concentration of objects close to the plane of the Local Supercluster extending to at least 230~Mpc. The effect is seen in the distributions of galaxies (in the 2MRS catalog\cite{2MRS} from the 2MASS infrared survey and in the independently obtained PSCz catalog\cite{PSCz} from the IRAS sky survey) and clusters of galaxies (identified by detections of X-ray emission by intracluster plasma\cite{clusters}). This seems quite unlikely to have grown out of Gaussian initial conditions. What seems more likely is that the cosmic initial conditions included seeds for the formation of these extended cosmic sheets that are  far thinner than their extent. I am led to the thought that reasonably straight cosmic strings moving through the universe after establishment of the Gaussian adiabatic departures from homogeneity would have left ridges of matter that grew into extended cosmic sheets such as the Local Supercluster.

\subsection{The merger tree}

The standard cosmology offers initial conditions and a physical theory that can be used to compute the  formation of cosmic structure---galaxies and their spatial distribution---that can be compared to what is observed in the universe as it is now and as it was at larger redshifts. The computations arrive at good approximations to the observations, which is important evidence that the standard cosmology is a useful approximation. The evidence is limited, however, by the complexity of the behaviour of the baryons, including the effects of formation and evolution of stars, black holes, and active galactic nuclei. That means apparent discrepancies between theory and observations must be treated with caution because it can be difficult to be sure of what the standard theory predicts. But I cannot resist commenting on a phenomenon that seems to be simply enough interpreted to be interesting for our purpose, the close to flat spiral galaxies.

A serious examination of this issue would start with a fair sample of galaxies whose properties are well enough known to allow discovery of the fraction that resemble the remarkably flat edge-on spirals seen nearby.  This is difficult at the present state of the art, but I think we can learn something of value from observations of examples among the closest galaxies that can be examined in greatest detail. I recommend looking at an image of the Silver Dollar Galaxy, officially NGC 253, which is readily found on the web. This galaxy is close enough, and it is tilted from edge-on just enough, to allow an inspection of its wonderfully close to flat surface. There is no obvious centre, no indication of the classical bulge of stars that is a customary part of a ``true'' spiral galaxy. The  Silver Dollar has a faint halo of stars spread over distances far greater than the size of the luminous part of the galaxy (illustrated in Fig.~1 in \cite{N253}). The visible stars in this halo amount to about 3\%\  of the stars in the Silver Dollar. You can find on the web images of other nearby spiral galaxies that happen to be seen edge-on, and are observed to be quite flat. Maybe these galaxies are not a fair sample, but they are real and their formation is an interesting challenge to standard thinking.

It is natural that ideas about how the galaxies formed are influenced by results of numerical simulations of galaxy formation. They suggest a merger tree: clumps of matter merged to form larger clumps that merged to form still larger clumps and so on in a hirerarchy of mergers that continued to the formation of galaxies and groups and clusters of galaxies as galaxy merging slowed. I offer as evidence of the influence of this concept the 51 papers that appeared in the single month of April 2024 and are found by the entry {\it full:"merger tree" year:2024} in the NASA astrophysics data system. A check of the first 15 indicates that at least 40 of the 51 papers appearing in April place substantial weight on the merger tree concept of galaxy formation. ADS returns 397 papers containing the phrase "merger tree" that appeared in the year 2023. Some will refer to other applications of a merger tree concept, but the evidence from the month of April is that most have to do with galaxy formation. The merger tree concept is influential.

My concern is that when two subhalos merge the stars that have already formed in them will end up in a stellar halo or bulge. One reads of the expected destruction of a growing disk by a merger that is followed by the formation of another disk. Again, the stars in the original disk will have joined the stellar bulge or halo of the growing galaxy. But this picture does not naturally agree with the properties of the Silver Dollar galaxy. It does not have a classical bulge of stars and only a few percent of its observed stars are in its extended stellar halo. From an empirical point of view the formation of this galaxy is better described by the classic Eggen, Sandage, and Lynden-Bell picture of the formation of the Milky Way galaxy by monolithic collapse\cite{EggenLynden-BelSandage}. So is the Silver Dollar galaxy a curious exception to the rule? Many edge-on and quite thin galaxies are seen nearby. Are these curious exceptions too? I do not know whether the nearby thin galaxies are true anomalies, but I do think it is awkward that these elegant objects require special pleading in the standard cosmology. 

\subsection{The two world pictures}

We have two well-tested fundamental theories of the nature of the world: particle physics on small scales and the $\Lambda$CDM cosmology on large scales. Particle physics passes many tests. It is incomplete, of course, but a better theory will have to include the many successful elements of the present theory. Cosmology is less thoroughly tested but I think well enough to expect that a better theory also will contain the empirically established elements of the $\Lambda$CDM theory.

We are accustomed to natural relations among physical theories. Consider that Maxwell's equations are the covariant vector theory with the simplest Lagrangian density that conserves charge, and Einstein's general theory of relativity is the covariant tensor theory with the simplest Lagrangian density that conserves local energy and momentum. This is an elegant natural relation. But it is difficult to see a natural relation between our two fundamental world pictures. On large scales we think of curved dynamical classical spacetime. On small scales we start with state vectors in an abstract space that allows entangled states that seem  ``spooky'' but are observed. 

The two fundamental theories give us four characteristic quantities: Hubble's constant $H_o$, Newton's constant $G$, Planck's constant $\hbar$, and the speed of light $c$. Hubble's constant gives us a characteristic expansion time, $H_o^{-1}\sim 10^{10}$~years, which is comparable to the astronomers' oldest stellar evolution ages. Cosmology and gravity physics, represented by Hubble's constant and Newton's constant, give us the characteristic mass density $H_o^2/G\sim 10^{-28}\hbox{ g cm}^{-3}$, which is a useful approximation to the mean mass density in stars. This was one of the reasons we paid attention to cosmology when I started looking into this physics in the 1960s. A theory that naturally produces a reasonable expansion time and cosmic mass density cannot be all bad. But quantum physics and gravity give us the  characteristic Planck mass $\sqrt{\hbar c/G}\sim 10^{-5}~\hbox{g}$ and Planck mass density $c^5/(\hbar G^2)\sim 10^{93}\hbox{ g cm}^{-3}$. (Apart from numerical factors this density is the familiar sum of zero-point energies of modes of oscillation of the electromagnetic field to the Planck length.) What are we to make of the Planck mass? Maybe some kinds of plankton have this mass but I suppose that could only be a coincidence. And what are we to make of the Planck mass density? In short, gravity physics and cosmology give naturally interpreted characteristic quantities, but gravity physics and quantum physics do not. 

For the most part our two fundamental theories peacefully coexist. At redshifts of galaxies that can be observed it is an excellent approximation to think of the quantum physics of atoms, molecules, and  masers operating in the given gently curved and adequately close to static classical spacetime of general relativity, thus operating very much as they do in the laboratory. And in astronomy and cosmology we need not think about the probabilistic nature of quantum physics; all that is averaged out in what astronomers observe. Particle physicists have wonderfully productive ideas drawn from quantum physics operating in Minkowski's flat static spacetime. They have not had to take account of the general expansion of the universe. 

I know of just one serious violation of coexistence, the absurdly large value of the characteristic quantum mass density compared to the value of Einstein's cosmological constant. Other challenges are to be expected in the physics of the very early stages of expansion of the universe, but that is work in progress. It might involve a version  of cosmological inflation, though this picture is at hazard of predicting an unacceptable excess of old universes in our past light cone. This is one of the ample supply of puzzles we are leaving for coming generations of physicists.

\subsection{The sky is not falling}

I could have mentioned other open issues in cosmology: the apparently small abundance of antimatter;  the physical natures of dark matter and dark energy; the origin of our universe, of physics, and of life; how the world will end; what is to be made of the multiverse; and what it all means. This is not ominous. It means there are many lines of enquiry that could enrich our physics. This is not unprecedented. All our physical theories are incomplete and active research aims to improve them. The rule, at least since Copernicus rationalized the Ptolemaic theory of the motions of the planets, is to preserve what is empirically successful. I see no reason to doubt that this will continue with future advances in cosmology. A theory of the world on large scales will be found and tested and judged to be a better approximation to reality than what we have now. And $\Lambda$CDM will be a recognizable aspect because it passes demanding empirical tests. 

Nonscientists tend to be particularly interested in discoveries in astronomy and cosmology, and we are responsible for communicating to the public a sensible impression of what we are doing. I dislike the tendency in the media and our own literature to represent open issues as crises or even failures. It attracts clicks. But in physical cosmology, with its solidly tested empirical basis that is not likely to go away, open issues are far more likely to be opportunities to learn more about the nature of physical reality.

\section{The cosmological principle}\label{sec:cos_prin}

The claim of a challenge to the foundational assumption of cosmology commonly appears in papers on the subject of Section~\ref{sec:dipole_anomaly} (in \cite{Sarkar_etal},\cite{Singal} and references therein). It sounds dramatic, a challenge to the basis for our subject. It is likely to disappoint those who pause to consider that the challenge depends on the definition; it could confuse others; and it distracts attention from the potentially important phenomenon discussed in Section~\ref{sec:dipole_anomaly}. 

Einstein argued that a philosophically satisfactory universe is homogeneous, apart from local fluctuations. The idea from simplicity was influential, though astronomers objected that the nearby galaxies are distributed in a decidedly inhomogeneous way. Hubble's \cite{Hubble} 1934 empirical approach was to compare counts of the faintest most distant galaxies detectable with the 100-inch telescope on Mount Wilson in directions spread across the accessible sky. It led to Hubble to conclude that 
\begin{quotation}
\noindent On the grand scale, however, the tendency to cluster averages out. The counts with large reflectors conform rather closely with the theory of sampling for a homogeneous population. Statistically uniform distribution of nebulae appears to be a general characteristic of the observable region as a whole.
\end{quotation} 
The theorist Robertson's \cite{Robertson} 1955 take was that the cosmological principle is
\begin{quotation}
\noindent the notion that the distribution and motion of matter in any sufficiently large spatial region of this universe are, by and large, intrinsically much the same as those in any other similar region, regardless of its position and orientation. This presumed uniformity in the large implies a certain form of a principle of relativity, sometimes called, appropriately enough, the ``cosmological principle.''
\end{quotation}
This is qualitative but appropriate. The earliest discussion along the present line that I have found began with the 1952 paper by the statisticians Jerzy Neyman and Elisabeth Scott\cite{NeymanScott} on a formal definition of what they  termed ``quasi-uniformity.'' Neyman \cite{Neyman} followed that in 1962 with the proposal that
\begin{quotation}
\noindent the universe, with all of its agglomerations of matter and the varied motions, is a single realization of a stochastic process [that is] \ldots stationary in the three spatial coordinates.
\end{quotation} 
This is well put, though I wish Neyman had added, ``stationary also in the two angular directions.''  

Another consideration is that in the standard cosmology cosmic structure grew by gravity out of departures of the primeval mass distribution from exact homogeneity. In the early universe the physical extents of the mass density fluctuations that grew into galaxies and groups and clusters of galaxies were far greater than the particle horizon computed subsequent to inflation, or whatever set initial conditions for the standard model. This is a violation of one form of the cosmological principle, isotropy on the largest observable scales\cite{Sarkar_etal},\cite{Singal}. And the assumed power law power spectrum of departures from homogeneity means the standard theory still is violating it.

The cosmological principle was foundational because it allowed research in cosmology to start with the evolution of the universe modelled as spatially homogeneous and isotropic. Interest in this first approximation was encouraged by the recognition that in a homogeneous and isotropic expanding universe the recession velocities of galaxies are proportional to their distances. In the 1930s Hubble and Humason confirmed the prediction for recession velocities up to a tenth of the speed of light. We later had the angular distribution of the CMB, with departures from isotropy so small they invited studies of their origin in linear perturbation theory. That led to the demanding tests mentioned in Section~\ref{sec:evidence}. A central challenge now is to better understand the nature and origin of departures from homogeneity in the distribution of matter in its various forms. 

The concept of statistical homogeneity and isotropy is an assumption, to be tested along with the other concepts of the standard cosmology. Milne (1933) elevated the assumption to ``Einstein’s cosmological principle.'' It made sense because there was so little other guidance to constructing a cosmology. That has changed, and I hope it is understood why I do not consider the assumption challenged by the present empirical evidence.

\section{Concluding thoughts}

This essay is written from an empiricist's point of view. Theories that go well beyond what can be connected to observations are to be respected; they can be elegant and prescient. Einstein's general theory of relativity was an accepted part of theoretical physics for a half century before tests began making the compelling case for it we now have. But the only way to be confident an idea, no matter how beautiful, has a secure contact with reality is from checks from the empirical evidence. I have not polled colleagues on this opinion, and I have encountered a few who might disagree. But I hold to it and to the belief that natural scientists do too, when they pause to think about it.

We can be sure research in progress will advance the physical sciences, including cosmology, by resolving troubling aspects of known issues and creating new ones. Much of this research requires large teams that must focus on specific goals. This is essential but it hinders recognition of other possibly interesting hints to the nature of reality. It at least in part accounts for the knowledge but lack of recognition of the extended plane of the Local Supercluster. There must be other examples of this sort. It is worth widening your vision on occasion to see if you can spot one, while keeping your day job.

\ack{I am grateful to Princeton University for allowing me space to try to widen my vision. Discussions with Professor Subir Sarkar led me to add section~\ref{sec:cos_prin}}.


\end{document}